\documentclass[showpacs,preprintnumbers, twocolumn]{revtex4-2}

\usepackage[colorlinks]{hyperref}

\usepackage[centertags]{amsmath}
\usepackage{amsfonts}
\usepackage{amssymb}
\usepackage{amsthm}
\usepackage{newlfont}
\usepackage{graphicx}
\usepackage{lipsum}
\usepackage{comment}

\usepackage{tikz}
\usetikzlibrary{calc}
\usepackage{tikz-3dplot}
\usetikzlibrary{patterns}
\usetikzlibrary{shadows}

\newtheorem{thm}{Theorem}[section]

\newtheorem{defn}[thm]{Definition}

\begin{document}

\title{Entanglement of temporal sections as quantum histories and their quantum correlation bounds }
\author{Marcin Nowakowski}
 \email[Electronic address: ]{marcin.nowakowski@pg.edu.pl}

\affiliation{Faculty of Applied Physics and Mathematics, Gdańsk University of Technology, 80-233 Gdańsk, Poland}
\affiliation{National Quantum Information Center, 80‑309 Gdańsk, Poland}


\begin{abstract}
In this paper we focus on the underlying quantum structure of temporal correlations and show their peculiar nature which differentiate them from spatial quantum correlations. 
With a growing interest in representation of quantum states as topological objects, we consider quantum history bundles based on the temporal manifold and show the source of violation of monogamous temporal Bell-like inequalities. We introduce definitions for the mixture of quantum histories and consider their entanglement as sections over the Hilbert vector bundles. As a generalization of temporal Bell-like inequalities, we derive the quantum bound for multi-time Bell-like inequalities.

\end{abstract}

\maketitle

\section{Introduction}

Recent years have witnessed increasing interest in the concept of quantum entanglement monogamy, demonstrating its utility in quantum communication theory and its applications in quantum secure key generation \cite{Lutk1, Lutk2, Lutk3, Lutk4, Devetak05, KLi}. While spatial quantum correlations, particularly their non-locality, have become a central focus of quantum information theory and its applications in quantum computation, the potential for applying temporal non-local correlations has been relatively underexplored. However, there is a growing interest in this area, which is linked to a better understanding of this distinctive quantum phenomenon.

A crucial issue relates to the fundamental nature of time and the phenomenon of temporal correlations, and their interpretation within the frameworks of modern quantum and relativistic theories. This emerging interest signifies a pivotal shift in focus, aiming to unravel the complexities of temporal phenomena in quantum mechanics and their implications for broader theoretical constructs.

Non-local nature of quantum correlations in space has been accepted as a consequence of violation of local realism (LR), expressed in Bell's theorem \cite{Bell} and analyzed in many experiments \cite{Aspect, Freedman}. As an analogy for a temporal domain, violation of macro-realism (MR) \cite{LGI2} and Leggett-Garg inequalities (LGI) \cite{LGI} seem to indicate non-local effects in time and are a subject of many experimental considerations \cite{Chu, EX1, EX2, EX3, EX4}. 
There have been different formalisms proposed for study of quantum temporal correlations including Multiple-Time States (MTS) by Aharonov {\it et al.} \cite{AAD,MTS1} as part of the Two-State-Vector formalism (TSVF) \cite{ABL,Properties,TSVFR,TTI}, the Entangled Histories (EH) approach  \cite{Cot} or the pseudo-density operators (PDOs) \cite{Vedral1, Marletto2, Zhang}.

The TSVF led to surprising effects within pre- and post-selected systems (e.g. \cite{Paradoxes,Pigeon,Dis}), time travel with post-selected teleportation \cite{Lloyd1,Lloyd2}, a novel notion of quantum time \cite{Each}, new results regarding quantum state tomography \cite{MTS2} and a better understanding of processes with indefinite causal order \cite{NewSandu}, while the Entangled Histories approach led to Bell tests for histories \cite{WC3} and have been recently used for analysis of the final state proposal in black holes \cite{CN}. The subject of the black hole information loss paradox has been also addressed with application of PDOs \cite{Marletto, Marletto3} but engaging a concept of non-monogamy of spatio-temporal correlations.   

In this paper we study the nature of the non-monogamous behavior of temporal correlations both for ensembles of quantum processes and their single instances. A central finding of our research, as highlighted in this letter, is the distinct nature of temporal versus spatial quantum correlations. In spatial quantum correlations, analysis typically involves ensembles composed of identical copies of multipartite states. Conversely, in the realm of temporal correlations, we encounter ensembles comprising diverse temporal histories. These histories serve as temporal counterparts to quantum states. A particular entangled history, which can be associated with a quantum propagator, is monogamous to  conserve  its  consistency  throughout  time \cite{Nowakowski}.   Yet  evolving  systems  violate  monogamous  Bell-like multi-time  inequalities which can be explained engaging bundles of histories with the same pre-selected and post-selected states as initial and final boundaries for the considered evolution. This dichotomy does not have a counterpart in spatial domain and as such is a novel feature of temporal non-locality but is also a sign of importance of the internal structure of single processes.

Our discussion shows the importance of considering the topological aspects of quantum processes, in addition to their statistical characteristics. Thus, we consider quantum histories as quantum vector 
bundles based on the temporal manifold and we introduce a definition for the mixture of quantum histories considering
their entanglement as sections over these vector bundles.

\section{Review of Entangled Histories and Multiple-Time States}

Let us review briefly the entangled histories (EH) formalism and the multiple-time states (MTS) formalism as a natural extension of the two-state vector formalism (TSVF). 

The predecessor of the entangled histories is the decoherent histories approach built on the grounds of the well-known Feynman’s path integral theory for calculation of probability amplitudes of quantum processes. The EH formalism extends the concepts of the consistent histories theory by allowing for complex superposition of histories. A history state is understood as an element in $\text{Proj}(\mathcal{H})$, spanned by projection operators from $\mathcal{H}$ to $\mathcal{H}$, where  $\mathcal{H}=\mathcal{H}_{t_{n}}\odot...\odot\mathcal{H}_{t_{0}}$.
The $\odot$ symbol, which we use to comply with the current literature, stands for sequential tensor products, and has the same meaning as the tensor product $\otimes$.
The alternatives at a given instance of time form an exhaustive orthogonal set of projectors  $\sum_{\alpha_{x}}P_{x}^{\alpha_{x}}=\mathbb{I}$ and for the sample space of entangled histories $|H^{\overline{\alpha}})=P_{n}^{\alpha_{n}}\odot P_{n-1}^{\alpha_{n-1}}\odot\ldots\odot P_{1}^{\alpha_{1}}\odot P_{0}^{\alpha_{0}}$ ($\overline{\alpha}=(\alpha_{n}, \alpha_{n-1},\ldots, \alpha_{0})$), there exist a set of $c_{\overline{\alpha}} \in \mathbb{C}$ such that $\sum_{\overline{\alpha}}c_{\overline{\alpha}}|H^{\overline{\alpha}})=\mathbb{I}$ and $\sum_{\overline{\alpha}} |c_{\overline{\alpha}}|^2=1$.

As an example, one can take a history $|H)=[z^+]\odot[x^-]\odot[y^-]\odot[x^+]=[|z^+\rangle\langle z^+|]\odot[|x^-\rangle\langle x^-|]\odot[|y^-\rangle\langle y^-|]\odot[|x^+\rangle\langle x^+|]$ for a spin-$\frac{1}{2}$ particle being in an eigenstate of the Pauli-X operator at time $t_1$, in an eigenstate of the Pauli-Y operator at time $t_2$, and so on. Within this formalism one also defines the unitary bridging operators $\mathcal{T}(t_j,t_i):\mathcal{H}_{t_i}\rightarrow\mathcal{H}_{t_j}$ evolving the states between instances of time, and having the following properties: $\mathcal{T}(t_j,t_i)=\mathcal{T}^{\dagger}(t_i,t_j)$ and $\mathcal{T}(t_j,t_i)=\mathcal{T}(t_j,t_{j-1})\mathcal{T}(t_{j-1},t_i)$.
This formalism introduces also the chain operator $K(|H^{\overline{\alpha}}))$, which can be directly associated with a time propagator of a given quantum process:
\begin{equation}
K(|H^{\overline{\alpha}}))=P_{n}^{\alpha_n}\mathcal{T}(t_{n},t_{n-1}) P_{n-1}^{\alpha_{n-1}}\ldots P_{1}^{\alpha_1}\mathcal{T}(t_{1},t_{0})P_{0}^{\alpha_0}
\end{equation}
This operator plays a fundamental role in measuring a weight of any history $|H^{\alpha})$:
\begin{equation}
W(|H^{\alpha}))=Tr\{K(|H^{\alpha}))^{\dagger}K(|H^{\alpha}))\}
\end{equation}
which can be interpreted as a realization probability of a history by the Born rule application. The histories approach requires also that the family of histories is consistent, i.e. one can associate with a union of histories a weight equal to the sum of weights
associated with particular histories included in the union.

Multiple-Time States (MTS) extend the standard quantum mechanical state by allowing its simultaneous description in several different moments. Such a multiple-time state may encompass both forward- and backward-evolving states on equal footing. MTS represent all instances of collapse (i.e. those moments in time when the quantum state coincided with an eigenstate of some measured operator) and allow them to evolve both forward and backward in time. This evolution backwards in time can be understood literally (giving rise to the Two-Time Interpretation \cite{TTI}), but this is not necessary, it can be simply regarded as a mathematical feature of the formalism (which is, in fact, equivalent to the standard quantum formalism \cite{TSVFR}). MTS live in a tensor product of Hilbert spaces $\mathcal{H}$ admissible at those various instances of time ($t_0<...<t_n$) denoted by \cite{MTS1}
\begin{equation}\label{HMTF}
\mathcal{H}=\mathcal{H}_{t_{n}}^{(\cdot)}\otimes...\otimes\mathcal{H}_{t_{k+1}}^{\dagger}\otimes\mathcal{H}_{t_{k}}\otimes\mathcal{H}_{t_{k-1}}^{\dagger}\otimes...\otimes\mathcal{H}_{t_{0}}^{(\cdot)},
\end{equation}
where a dagger means the corresponding Hilbert space consists of states which evolve backwards in time. The initial and final Hilbert spaces might be daggered or not (this is denoted by a ``$\cdot$'' superscript). All Hilbert spaces containing either (forward-evolving) kets or (backward-evolving) bras are alternating to allow a time-symmetric description at any intermediate moment. 

As an example of (a separable) MTS we can consider the following state: $_{t_4}\langle z^+||x^-\rangle_{t_3~t_2}\langle y^-||x^+\rangle_{t_1} \in \mathcal{H}_{t_{4}}^\dagger\otimes\mathcal{H}_{t_{3}}\otimes\mathcal{H}_{t_{2}}^\dagger\otimes\mathcal{H}_{t_{1}}$. This multiple-time state represents an initial eigenstate of the Pauli-X operator evolving forward in time from $t_1$ until collapse into an eigenstate of the Pauli-Y operator occurs at time $t_2$. Later on, at time $t_3$ the system is projected again onto a different eigenstate of the Pauli-X operator. Finally at $t_4$ the system is measured in the Z basis, and the resulting eigenstate evolves backward in time. In the following we will focus on two-time states (sometimes called two-states), which consist of a forward evolving state $|\psi_1\rangle_{t_{1}}$ and a backward evolving state $|\psi_2\rangle_{t_{2}}$ in the above form $_{t_2}\langle \psi_2| |\psi_1\rangle_{t_1}$ to achieve a richer description of a quantum system during the time interval $t_1\le t \le t_2$ \cite{TSVFR}. 

Given an initial state $|\Psi\rangle$ and a final state $\langle \Phi|$, the probability that an intermediate measurement of some hermitian operator $A$ will result in the eigenvalue $a_n$ is given by the ABL formula \cite{ABL}

\begin{equation} \label{ABL}
p(A=a_n)=\frac{1}{N}|\langle\Phi|U_{2}P_{n}U_{1}|\Psi\rangle|^2,
\end{equation}
where $U_1$ and $U_2$ represent unitary evolution, the operator $P_n$ projects on $|a_n\rangle$ and
\begin{equation}
N \equiv \sum_k |\langle\Phi|U_{2}P_{k}U_{1}|\Psi\rangle|^2.
\end{equation}
This probability rule is important in that it uses the information available through the final state in a way which is manifestly time-symmetric.

\section{Bundles of quantum histories and their mixtures }

In this section we propose the mathematical representation of entangled quantum histories framed within the context of vector bundles. This approach offers a topologically structured way to understand the evolution of quantum states. Recently, there is a growing attention paid to these connections \cite{Hastings, Schreiber} trying to understand the role that vector bundles of spaces of quantum
states play in the classification of topological phases of matter and topological quantum computation via understanding how entangled structures occur in parametrized vector bundles.

 It has been also argued \cite{Codes} that quantum
codes can be described by section of a fiber bundle, where the base corresponds to a choice of
stabilizers of the code and the fiber describes the encoded logical information. Our motivation behind this path stems from the necessity of consideration of both particular quantum histories but also their mixtures which have substantial consequences for statements about their entropic characteristics. Topological representation and visualization of quantum histories as fibers and vector bundles allows us to capture the subtlety of violating Bell's temporal monogamous inequalities which is not obvious in case of TSVF or pseudo-density formalisms.
We delve now into more formal definitions and theorems to deepen this understanding.

A vector bundle $\mathcal{E}$ over a base  space $\mathcal{M}$ is a topological structure associating family of vector spaces $\mathcal{E}(m)$ to elements of space $\mathcal{M}$.

We present now two representations of quantum histories as sections over vector bundles and fibers of vector bundles. In the first representation the vector bundle is built upon a temporal manifold $T$ and a particular history is a consistent section over the bundle. In the second representation it is indexed by a sequence of chosen measurements and quantum histories are fibers of the vector bundle.

Consider a manifold \( T \) representing time, each point in 
\( T\) corresponds to a distinct time in the evolution of the quantum system. Let \( \mathcal{H} \) denote a Hilbert space associated with our quantum system. For each time point 
t, there exists a Hilbert space $\mathcal{H}_{t}$ that is isomorphic to 
$\mathcal{H}\cong\mathcal{H}_{t}$. We define a vector bundle $\mathcal{E}$ over \( T \) where for each point \( t \in T \)  the fiber $\mathcal{E}_t\cong \mathcal{H}_t$,  represents the state space over that time.

A quantum history $|H_s)$ in this framework is a consistent section of the bundle $\mathcal{E}$. Formally, a section \( s: T \rightarrow \mathcal{E} \) is a map that assigns to each point in time \( t \) a state $\Psi(t)$ in the fiber \( \mathcal{E}_t \), which is a state of the quantum system at that time. This can be expressed as:

\begin{defn}
Let  $\mathcal{H}\ni\Psi(t)$ denotes a Hilbert space of possible states associated with a quantum system evolving over time $t\in T$. A history bundle $\mathcal{E}$ over $T$ associates with each measurable time $t$ the fiber $\mathcal{E}_t\cong \mathcal{H}_t$ so that the evolution of the system is represented by a consistent section $s=|H_s)$ over that bundle, i.e.:

\begin{align}
s: T &\rightarrow \mathcal{E} \\
t &\mapsto s(t)
\end{align}
where \( s(t) \in \mathcal{E}_t \) for each \( t \in T \).

\end{defn}

Histories are entangled if their corresponding sections do not factorize into independent states across different times.
This representation allows us to view quantum histories as continuous (or piece-wise continuous) trajectories in the state space of the system. It emphasizes the temporal evolution of quantum states, capturing the essence of quantum dynamics over time. This reasoning can be naturally extended to the base space-time manifold $\mathcal{M}(x,t)$ and space of field histories $\Phi(x,t)$. 

Now we propose a second less intuitive representation where quantum histories are fibers of vector bundles over the measurement sequences space.
Let us consider the base space $\mathcal{M}$ of all possible finite sequences of measurements, where $\mathcal{M}= \bigcup_{n=1}^{\infty} M^n $, and \( M \) is the set of all possible measurements. A vector bundle \( F \) is constructed over \( \mathcal{M} \) where each fiber \( F_\alpha \), for a measurement sequence \( \alpha \in \mathcal{M} \), represents the possible outcomes of that sequence.
Recalling that the alternatives at a given instance of time form an exhaustive orthogonal set of projectors  $\sum_{\alpha_{x}}P_{x}^{\alpha_{x}}=\mathbb{I}$ and for the sample space of entangled histories $|H^{\overline{\alpha}})=P_{n}^{\alpha_{n}}\odot P_{n-1}^{\alpha_{n-1}}\odot\ldots\odot P_{1}^{\alpha_{1}}\odot P_{0}^{\alpha_{0}}$ ($\overline{\alpha}=(\alpha_{n}, \alpha_{n-1},\ldots, \alpha_{0})$), there exist a set of $c_{\overline{\alpha}} \in \mathbb{C}$ such that $\sum_{\overline{\alpha}}c_{\overline{\alpha}}|H^{\overline{\alpha}})=\mathbb{I}$ and $\sum_{\overline{\alpha}} |c_{\overline{\alpha}}|^2=1$, one can construct a morphism $\overline{\alpha}\rightarrow |H^{\overline{\alpha}})$ which is a history fiber over the sequence of projective operators.

In this approach, each fiber \( F_\alpha \) is a chain of states that the system can reach following the sequence \( \alpha \) from an initial pre-selected state.

With this formal divagation about the entangled histories, field histories and their connection with observability, readers are invited to reach out to the theory of an internal observer \cite{Nowakowsk2} proposing hierarchical observability and emphasizing profound role of observers in creation of reality. These topological studies will be extended in the paper \cite{NowXerxes} about homotopy spaces and their applications to the theory of higher-order quantum histories.

The fundamental property of multipartite spatial quantum entanglement is its monogamy. This property states that for the case of tripartite system ABC, maximal entanglement of the pair AB excludes its entanglement with the third party, i.e. if $\rho_{AB}=|\Psi^+\rangle\langle\Psi^+|$, then any extension of this state is of the form $\rho_{ABC}=|\Psi^+\rangle\langle\Psi^+|\otimes |\Psi\rangle\langle\Psi|$. For the temporal correlations, it seems that this property does not hold, especially when one considers statistical distribution of measurement results \cite{White, Marletto}. Yet, what is obvious in the spatial case does not have mere analogies in the temporal case. It was proved \cite{Nowakowski} that a particular history can be monogamous but further we will discuss how temporal correlations can lead to non-monogamous results for bundles of histories with which we tackle during the measurement process. This subtlety is rather a sign of a deeper nature of quantum processes which can keep their consistency for particular instances, yet leads to quite counter-intuitive results for their ensembles.  

Suppose we have two non-equivalent multi-time entangled histories of an evolving qubit through times ${t_4> t_3> t_2> t_1}$ for which we consider the past effect of the measurement at time $t_4$: 

\begin{eqnarray}
    |H_1)&=&\frac{1}{\sqrt{2}}[|0)\odot|0)\odot|0)\odot|0)+|1)\odot|1)\odot|1)\odot|1)] \nonumber \\
    |H_2)&=&\frac{1}{\sqrt{2}}|0)\odot[|0)\odot|0)\odot|0)+|1)\odot|1)\odot|1)]
\end{eqnarray}

The history $|H_1)$ can be perceived as a superposition of two histories on times ${t_4, t_3, t_2, t_1}$. If one measures this evolution at time $t_4$ with dichotomic projective observables $P_0=|0\rangle\langle 0|$ and $P_1=|1\rangle\langle 1|$, we can conclude that the state was with probability $p_0=\frac{1}{2}$ in a history $|H_{10})=|0)\odot|0)\odot|0)$ at previous times and with probability $p_1=\frac{1}{2}$ in a history $|H_{11})=|1)\odot|1)\odot|1)$. Alternatively, one can consider an ensemble of history states $\{\{p_0, |H_{10})\},\{p_1, |H_{11})\}\}$, i.e. half of the qubits evolving trivially in a history $|H_{10})$ and half in $|H_{11})$ through times ${t_3, t_2, t_1}$ which can be represented by a history super-operator $\rho_H=\frac{1}{2}(|H_{10})(H_{10}| + |H_{11})(H_{11}|)$. 
This evolution is different for the history $|H_2)$. If one performs the same measurements at time $t_4$, then we get an entangled history through times ${t_3, t_2, t_1}$ for the projective measurement $P_0$ at time $t_4$. Thus, physically we can propose the concept of \textit{the probabilistic mixture of histories}:

\begin{defn}
A mixed history state is defined as a positive super-operator acting on a history state space:
\begin{equation}
\rho_{hist}=\sum_{i}p_i |H_i)(H_i|
\end{equation}
where $Tr\rho_{hist}=1$, $\sum_{i} p_i=1$ and $\forall_i 1>p_i\geq 0 $.
\end{defn}
This mixture of histories can be naturally associated with an ensemble of histories $\{p_i, |H_i)\}$ that can be understood as a mixture of sections over the history bundles.
Following, we consider an example of a spin particle traversing two paths to check a future influence of the measurement at time $t_1$:

\textit{Example 1. }Imagine a spin-$\frac{1}{2}$ particle at three times $\{t_3, t_2, t_1\}$ evolving trivially by $\mathcal{T}=\mathbb{I}$ with a family of entangled histories:
\begin{eqnarray}
|H^1)&=&\sqrt{2}([z^{+}]\odot[x^{+}]\odot[z^{+}]+[z^{-}]\odot[x^{-}]\odot[z^{+}])\nonumber\\
|H^2)&=&\sqrt{2}([z^{-}]\odot[x^{+}]\odot[z^{+}]+[z^{+}]\odot[x^{-}]\odot[z^{+}])\nonumber\\
|H^3)&=&\sqrt{2}([z^{+}]\odot[x^{+}]\odot[z^{-}]+[z^{-}]\odot[x^{-}]\odot[z^{-}])\nonumber\\
|H^4)&=&\sqrt{2}([z^{-}]\odot[x^{+}]\odot[z^{-}]+[z^{+}]\odot[x^{-}]\odot[z^{-}])\nonumber\\
\end{eqnarray}
If we consider a state $|\Phi)=\frac{1}{\sqrt{2}}|H^1)+\frac{1}{\sqrt{2}}|H^2)$, then a particle, measured at time $t_1$ and having a spin up in a direction $z^{+}$, can evolve within the history $|H^1)$ with probability $P(|H^1))=\frac{1}{2}$ and be in the history $|H^2)$ with probability $P(|H^2))=\frac{1}{2}$. It is also interesting to observe that $|\Phi)=\frac{1}{\sqrt{2}}|H^1)+\frac{1}{\sqrt{2}}|H^2)=[z^{+}]\odot[z^{+}]\odot[z^{+}]$ where $[z^{+}]=[|z^{+}\rangle\langle z^{+}|]$.\\
Noteworthily, one can also find in the space of histories $\mathcal{S}=span\{|H^1), |H^2), |H^3), |H^4)\}$ the following temporal GHZ-like vector \cite{Cot} (normalized for $|\alpha|^2+|\beta|^2=1$):
\begin{eqnarray}
|\tau GHZ)&=&\frac{\alpha}{\sqrt{2}}|H^1)+\frac{\alpha}{\sqrt{2}}|H^2)+\frac{\beta}{\sqrt{2}}|H^3)+\frac{\beta}{\sqrt{2}}|H^4)\nonumber\\
&=&\alpha[z^{+}]\odot[z^{+}]\odot[z^{+}]+\beta[z^{-}]\odot[z^{-}]\odot[z^{-}] \nonumber\\
\end{eqnarray}


Let us consider now a quantum history displaying temporal entanglement at times $\{t_2, t_1\}$:
\begin{equation}
    |H)=\alpha|\phi_{3,2})\odot(|\phi_{2,1})\odot |\phi_{1,1})+|\phi_{2,2})\odot |\phi_{1,2}))\odot |\phi_0)
\end{equation}
This particular history can be realized by placement of a detector at time $t_3$ which detects a state $\rho=|\phi_{3,2}\rangle\langle \phi_{3,2}|$ and displays quantum entanglement in time for times $\{t_2, t_1\}$: 
\begin{equation}
 |H_{t_2, t_1})=\alpha(|\phi_{2,1})\odot |\phi_{1,1})+|\phi_{2,2})\odot |\phi_{1,2}))   
\end{equation}
Interestingly, this entangled history at times $\{t_2, t_1\}$ cannot be derived from the following temporal version of a GHZ-state manifesting monogamy of temporal entanglement for particular histories \cite{Nowakowski}:
\begin{equation}
|\widetilde{H})=\alpha(|\phi_{3,1})\odot |\phi_{2,1})\odot |\phi_{1,1})+|\phi_{3,2})\odot |\phi_{2,2})\odot |\phi_{1,2}))\odot |\phi_0)
\end{equation}

Let us observe that the reduced component of this history $|\phi_{3,1})\odot |\phi_{1,1})$ is correlated with $|\phi_{2,1})$ and not with $|\phi_{2,2})$. Thus, reduction of $|\widetilde{H})$ over times $t_2$ and $t_0$ is not a complex superposition of histories but is \textit{a probabilistic mixture} as already stated in this section:
\begin{equation}
\rho_{t_1t_3}=|\alpha|^2(|\phi_{3,1}\phi_{1,1})(\phi_{3,1}\phi_{1,1}|+|\phi_{3,2}\phi_{1,2})(\phi_{3,2}\phi_{1,2}|)    
\end{equation}
where $|\phi_{3,1}\phi_{1,1})=|\phi_{3,1})\odot |\phi_{1,1})$ etc.
This can be also formally derived employing a temporal  partial trace operator \cite{TraceTime} over time instances:
$\rho_{t_1t_3}=Tr_{t_2t_0}|\widetilde{H})(\widetilde{H}|$. This operator is analogous to spatial tracing out but must maintain consistency in the evolution—a condition that is not present in the spatial case.

\section{What entangled histories say about temporal bounds}

The violation of local realism \cite{Bell} and macrorealism \cite{LGI2} by quantum theories has been studied for many years in experimental setups where measurement data is tested against Bell inequalities for LR and Leggett-Garg inequalities \cite{LGI} for MR. In quantum theories, the former arises as a consequence of non-classical correlations in space, while the latter arises as a consequence of non-classicality in dynamic evolution.

In the temporal version of the CHSH-inequality (for Clauser, Horne, Shimony, and Holt \cite{CHSH}) being a modification of original Leggett-Garg inequalities, Alice performs measurement at time $t_{1}$ choosing between two dichotomic observables $\{A_{1}^{(1)}, A_{2}^{(1)}\}$ and then Bob performs a measurement at time $t_{2}$ choosing between $\{B_{1}^{(2)}, B_{2}^{(2)}\}$. Therefore, the structure of this LGI can be represented as follows \cite{Vedral}:
\begin{equation}
S_{AB}\equiv c_{12}+c_{21}+c_{11}-c_{22} \leq 2
\end{equation}
where $c_{ij}=\langle A_{i}^{(1)}, B_{j}^{(2)}  \rangle$ stands for the expectation value of consecutive measurements performed at time $t_{1}$ and $t_{2}$.

The general LGI for n-time consecutive measurements is given by:
\begin{equation}
K_n\equiv c_{12}+c_{23}+\ldots+c_{n-1n}-c_{1n}
\end{equation}
bounded for odd n as $-n\leq K_n\leq n-2$ and for even n as $-(n-2)\leq K_n\leq n-2$. These inequalities are naturally violated in quantum realm and for qubits they reach so-called L\"{u}ders bounds, i.e. $K_{nQ}=ncos(\frac{\pi}{n})$ \cite{Budroni1, Budroni2}.

Since one can build in a natural way $\mathcal{C}^{*}$-Algebra of history operators for normalized histories from projective Hilbert spaces equipped with a well-defined inner product, the quantum Tsirelson bound $2\sqrt 2$ of CHSH-inequality specific for spatial correlations holds also for temporal LGI engaging only the space of entangled histories \cite{Nowakowski}.

Let us consider a temporal setup with measurements $\mathbb{A}=I \odot \mathbb{A}^{(1)}$ (measurement $\mathbb{A}$ occurring at time $t_{1}$) and $\mathbb{B}=\mathbb{B}^{(2)}\odot I$. The history with 'injected' measurements can be represented as $|\widetilde{H})=\alpha \mathbb{A}\mathbb{B}|H)\mathbb{A}^{\dagger}\mathbb{B}^{\dagger}$ where $\alpha$ stands for a normalization factor.
History observables are history state operators which are naturally Hermitian and their eigenvectors can generate a consistent history family\cite{Cot}. As an example, we can consider  spin $\frac{1}{2}$-particle with a history inducing evolution $|\psi(t_1)\rangle\rightarrow|\psi(t_2)\rangle$ on which we act with $\sigma_y\odot\sigma_x$ operation. This step results with a new effective history:
\begin{equation}
|\widetilde{H})=\alpha \sigma_y[\psi(t_2)]\sigma_y^{\dagger} \odot \sigma_x[\psi(t_1)]\sigma_x^{\dagger}
\end{equation}
For an observable $A=\sum_{i}a_{i}|H_{i})(H_{i}|$, its measurement on a history $|H)$ generates an expectation value $\langle A\rangle=Tr(A|H)(H|)$ (i.e. the result $a_{i}$ is achieved with probability $|(H|H_{i})|^{2}$) in analogy to the spatial case. Thus, one achieves history $|\widetilde{H})$ as a realized
history with measurements and the expectation value of the history observable $\langle A \rangle$. It is worth mentioning that $|\widetilde{H})$ and $|H)$ are both compatible histories, i.e. related by a linear transformation. 

For temporal correlations measurements can lead to counter-intuitive results which do not occur for spatial quantum resources. Let us reexamine the case of $GHZ$ states firstly shared as a spatial system of three entangled qubits among Alice, Bob and Charlie: $|\Psi_{ABC}\rangle=\frac{1}{\sqrt{2}}(|000\rangle + |111\rangle)$ which obviously leads to a separable state for any pair from this system, e.g. $\rho_{AB}=\frac{1}{2}(|00\rangle\langle 00|+ |11\rangle\langle 11|)$. Assume further that they can choose from dichotomic projective observables: $P_0=|0\rangle\langle 0|$ or $P_1=|1\rangle\langle 1|$, then in this multipartite case any pair cannot identify alone without the third party that they are part of the more complex entangled system. In the temporal case, the situation is quite the opposite when considering the temporal version of the GHZ-state, highlighting a qualitative distinction between spatial and temporal resources. Alice, Bob, and Charlie, each possessing instances of the same system but at different points in time, can independently detect temporal non-locality within each pair.

When we measure an average value of the aforementioned Bell-like temporal inequality:
\begin{equation}
\langle S_{\tau AB} \rangle=\langle A_{1}B_{1}+A_{1}B_{2}+A_{2}B_{1}-A_{2}B_{2} \rangle
\end{equation}
we consider an ensemble of systems from which each quarter is measured against the observables $A_{i}B_{j}$.  It is easy to observe that with a choice of observables:
$A_1=Z, A_2=(Z+X)/\sqrt{2}, B_1=Z, B_2=(Z-X)/\sqrt{2}$ we get effectively the average value: $\langle S_{\tau AB} \rangle= \sqrt{2}\langle XX+ZZ \rangle$. Since one gets $\langle XX \rangle =\langle ZZ \rangle = 1$, the Tsirelson maximum is saturated. However, what is important in this simple example is that consecutive measurements of both X and Z leave the system in the same eigenstate for any number of time steps. As an immediate implications, one gets violation of monogamous Bell-like inequalities in space \cite{Pawlowski}:     
\begin{equation}
    S_{AB}+S_{BC}\nleq 4
\end{equation}
since for the temporal tripartite system $ABC$ (B and C being instances of A at consecutive times) we get saturation for the AB pair and for the BC:
\begin{equation}
S_{\tau AB}+S_{\tau BC}=4\sqrt{2}    
\end{equation}
This limit cannot be achieved through spatial correlations \cite{Pawlowski}. In spatial correlations, monogamy relations between the strengths of violations of Bell's inequalities are derived from the no-signaling condition. However, when it comes to temporal correlations, the situation differs because temporal correlations signal towards the future, causing the no-signaling condition to be violated, which, in turn, results in this inequality violation.

The fundamental aspect of generating such averaged Bell-like inequalities lies in our operation with a bundle of \textit{different} histories. These histories are steered by various collections of measurements, all beginning with the same initial pre-selected state and concluding with the same post-selected final state for the entire bundle. However, these histories involve different intermediary steps, as illustrated in the example with XXX and ZZZ quantum operations mentioned above. 
It was also observed in \cite{Vedral} that in the case of a multi-point temporal correlation function,
for measurements performed at m instances of time $t_1, \ldots, t_m$, the function
is decomposable into a product of two-fold temporal correlations. However, the temporal
correlation function is understood as the average over many runs of the sequence
of measurements. Thus, this result applies again to bundles of different histories.

We can look at the problem of bounding temporal correlations also by prism of the two-state vector  formalism which is isomorphic to the entangled histories \cite{NowakowskiCohen}. The correlations can be described by the probabilistic boxes in non-signalling theory. The box is shared between two parties who give the input setting $\{x, y\}$ of the measuring devices and get the outputs $\{a, b\}$ with probability $p(ab|xy)$ being an entry of the join probability distribution matrix $P(ab|xy)=[p(ab|xy)]$. All entries of this matrix meet the non-negativity condition ($p(ab|xy)\geq 0$) and are normalized:$\forall_{x,y}\sum_{a,b}p(ab|xy)=1$ and the no-signalling condition imposed on the quantum correlations by the special relativity constraints: the marginals $p(a|x)$ and $p(b|y)$ are independent of settings y and x respectively, i.e. $\forall_{y,a,x}p(a|x)=\sum_b p(ab|xy)$ and $\forall_{x,b,y}p(b|y)=\sum_a p(ab|xy)$.     
Then the Aharanov-Bergmann-Lebowitz (ABL) formula (\ref{ABL})  delivers a method for calculation of the measurements probability in between the initial time with the pre-selected state and the post-selected state at the final time of the analyzed quantum process.

\begin{figure}[h]
\centerline{\includegraphics[width=9.5cm]{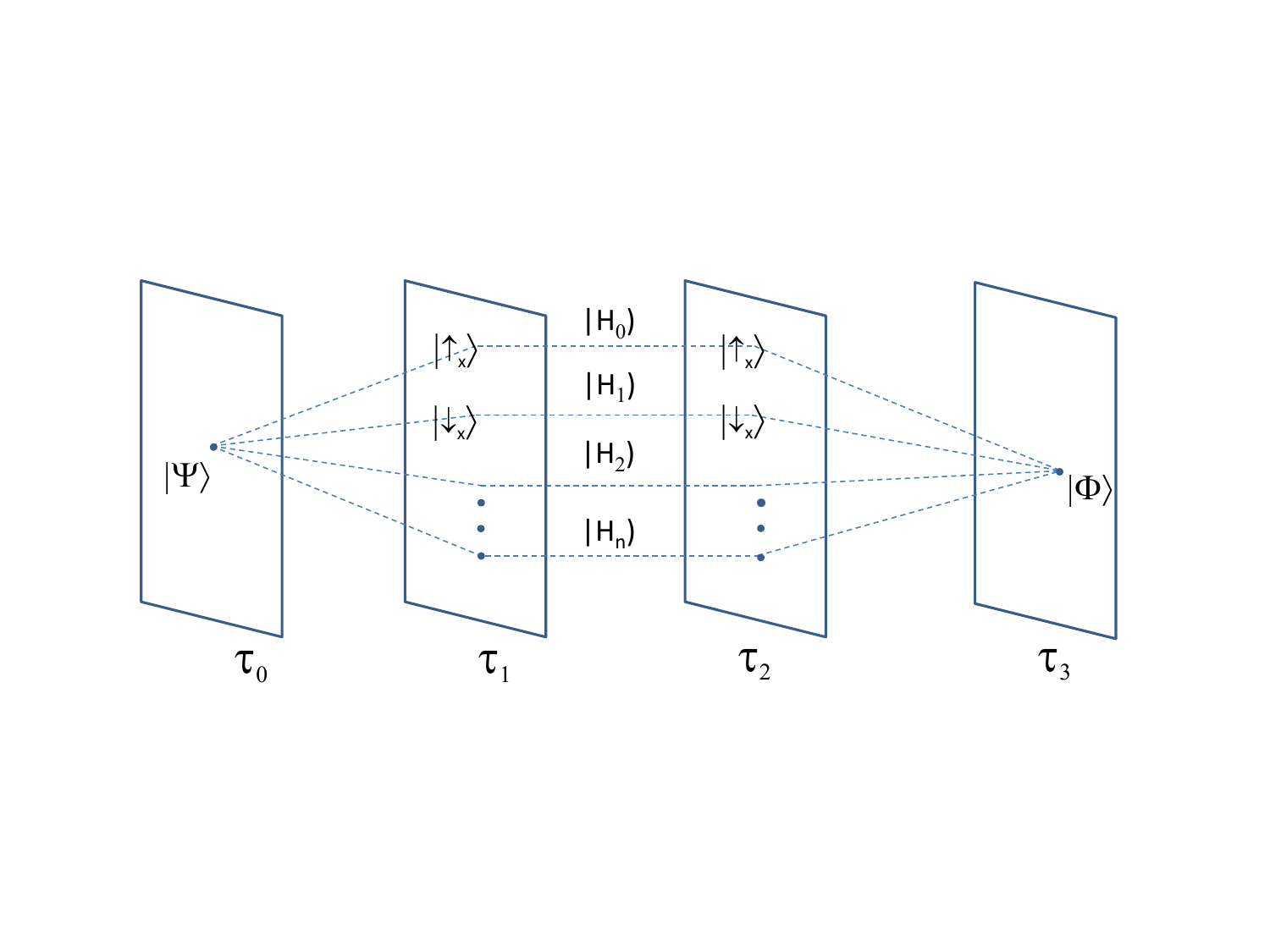}}
 \caption[GlobalHistory]{A bundle o histories at times ${\tau_0, \tau_1, \tau_2, \tau_3}$ with a pre-selected state $|\Psi\rangle$ and post-selected state $|\Phi\rangle$. Exemplary histories $|H_0)$ and $|H_1)$ with incorporated measurement results of X. This bundle contributes to violation of monogamous temporal Bell-like inequalities engaging different history states with the same initial and final states.}
 \label{GlobalHistory}
\end{figure}

In the case of series of X and Z measurements injected in the histories considered in this section we get the following example of probability distribution with assumption that at times $t_1$ and $t_2$ the X observable is chosen and we get $|\uparrow_x\rangle$ results in both times:  
\begin{equation}
p(\uparrow_x\uparrow_x|XX)=\frac{|\langle \Phi |\uparrow_x\rangle\langle\uparrow_x| |\uparrow_x\rangle\langle\uparrow_x| \Psi\rangle|^2}{\sum_{ab}p(ab|XX)}
\end{equation}
This is an operational method for generation of the whole probability distribution matrix. However, we should note that these experiments start with the same initial and final states but with different intermediate steps, thus, leading to a bundle of histories at times ${t_0, t_1, t_2, t_3}$ (Fig. \ref{GlobalHistory}):
\begin{equation}
    |H_{abxy}) \sim p(ab|XY)
\end{equation}
which can lead to violation of spatial quantum bounds on Bell-like inequalities.

We can then formulate generic bounds on temporal correlations of qubits in quantum theories (this result can be generalized to the qudits' case).

Let us assume that the quantum process occurs n times and that for any two times $\{t_i, t_{i+1}\}$ the quantum bound $\mathcal{Q}$ limits the temporal Bell-like functional on the matrix of probability distributions, i.e. $B_{\tau}(A_i,A_{i+1})=\mathcal{F}([p_{t_i, t_{i+1}}(ab|xy)])\leq \mathcal{Q}$ with the association of histories $|H_{abxy})$, then the process saturating the chain for such n-steps can be designed in such a way that each pair of times is a replication of two consecutive times, i.e. $\forall_i [p_{t_i, t_{i+1}}]=[p_{t_{i+2}, t_{i+3}}]$. This process is equivalent logically to a loop $t_0\rightarrow t_1 \rightarrow t_0\ldots\rightarrow t_1$. In consequence, we get the following quantum bound :

\begin{equation}
\sum_{i=0}^{n-1} B_{\tau}(A_i,A_{i+1})\leq \mathcal{Q}n
\end{equation}
that can be saturated to its maximal value.
As an implication for the LGIs one gets the following quantum bound:
\begin{equation}
    \sum_{i=1}^n S_{LGI}(A_0,A_i)\leq 2\sqrt{2}n
\end{equation}
which can be saturated to the maximal value in quantum world and which violates the spatial monogamy relations $ \sum_{i=1}^n B(A_0,A_i)\leq 2n$ (for $n\geq 2$) \cite{Pawlowski}.

We can conclude this section with a remark that a particular entangled history is monogamous, in similarity to quantum spatial states, but for a bundle of different histories with the same pre-selected and post-selected states one can get violation of monogamy. This is a novel feature of temporal correlations not paralleled in spatial domain.

\section{Conclusions}


In this study, we have ventured into the burgeoning domain of quantum state representation using topological paradigms, specifically focusing on quantum history bundles constructed upon temporal manifolds. This approach has allowed us to shed light on the mechanism that leads to the violation of monogamous temporal Bell-like inequalities, a topic that has garnered considerable attention in recent research.

With a growing interest in representation of quantum
states as topological objects, we consider quantum history bundles based on the temporal manifold and show
the source of violation of monogamous temporal Bell-
like inequalities. We introduce definitions for the mixture
of quantum histories and consider their entanglement as
sections over the Hilbert vector bundles. Our analysis reveals that, in contrast to spatial quantum correlations where ensembles consist of identical copies of multipartite states, temporal correlations involve ensembles of varying temporal histories. This fundamental distinction underscores a qualitative difference in the results obtained for spatial and temporal Bell-like inequalities, marking a significant advancement in our understanding of quantum phenomena.

Further advancing our theoretical understanding, we have utilized the Tsirelson bound to derive a quantum limit for multi-time Bell-like inequalities. This derivation, based on the entangled histories approach, offers a novel perspective on temporal quantum bounds.

The field of quantum state representation through topological methods presents numerous opportunities for further exploration. Specific avenues for future research could include the practical applications in quantum computing and quantum information theory. 

In conclusion, our study not only provides a deeper understanding of the temporal aspects of quantum states but also highlights the importance of considering topological, alongside statistical, aspects of quantum processes. The insights gained from this research pave the way for a more comprehensive and nuanced understanding of quantum mechanics, and we anticipate that they will stimulate further research in this exciting and rapidly evolving field.

\section{Acknowledgments}
Acknowledgments to Xerxes Arsiwalla, David Chester, Eliahu Cohen, Marek Czachor and Pawel Horodecki for discussions about temporal correlations.


\begin{thebibliography}{8}
\bibitem{TraceTime}
For a history $|Y_{t_{n}\ldots t_{0}})=F_{n}\odot F_{n-1}\odot\ldots\odot F_{1}\odot F_{0}$ acting on a space $\mathcal{H}=\mathcal{H}_{t_{n}}\otimes\dots\otimes\mathcal{H}_{t_{0}}$, a partial trace over times $\{t_{j}\ldots t_{i+1} t_{i}\}$ $(j\geq i)$ is:
$
Tr_{t_{j}\ldots t_{i+1}t_{i}} |Y_{t_{n}\ldots t_{0}})(Y_{t_{n}\ldots t_{0}}|=\sum_{k=1}^{\dim\mathcal{F}} (e_{k}|Y_{t_{n}\ldots t_{0}})(Y_{t_{n}\ldots t_{0}}|e_{k})
$
where $\mathcal{F}=\{|e_{k})\}$ creates an orthonormal consistent family of histories on times $\{t_{j}\ldots t_{i+1} t_{i}\}$ and the strong consistency condition for partial histories holds for base histories, i.e. $(e_{i}|e_{j})=Tr[K(|e_{i}))^{\dag}K(|e_{j}))]=\delta_{ij}$.
\end{thebibliography}
\end{document}